\documentclass[aps,pre,twocolumn,floats,floatfix,english,showpacs]{revtex4}
\usepackage[T1]{fontenc}
\usepackage[latin1]{inputenc}
\usepackage{amsmath}
\usepackage{graphicx}
\usepackage{setspace}
\usepackage{epsfig}
\usepackage{pstricks}
\usepackage{pst-node}
\usepackage{pst-plot}
\usepackage{latexsym}
\makeatletter

\providecommand{\LyX}{L\kern-.1667em\lower.25em\hbox{Y}\kern-.125emX\@}

\usepackage{babel}
\makeatother
\begin{document}

\title{Intensity and coherence of motifs in weighted complex networks}

\author{Jukka-Pekka Onnela$^1$}
%\email{jonnela@lce.hut.fi}
\author{Jari Saram\"aki$^1$}
\author{J\'anos Kert\'esz$^{1,2}$}
%\altaffiliation[Permanent address: ]{Department of Theoretical Physics,
%  Budapest University of Technology and Economics, Budapest, H-1111}
\author{Kimmo Kaski$^1$}
\affiliation{$^1$Laboratory of Computational Engineering,
  Helsinki University of Technology, Espoo, Finland\\
  $^2$Department of Theoretical Physics, Budapest University of Technology
  and Economics, Budapest, Hungary
}

%\date{\today}

\begin{abstract}
The local structure of unweighted networks can be characterized by the number of times 
a subgraph appears in the network. The clustering coefficient, reflecting on the 
local configuration of triangles, can be seen as a special case of this approach. In this Letter 
we generalize this method for weighted networks.  
We introduce subgraph \emph{intensity} as the geometric mean of its link weights and 
\emph{coherence} as the ratio of the geometric to the corresponding arithmetic mean. 
Using these measures,  motif scores and clustering coefficient can be generalized to weighted networks. 
To demonstrate these concepts, 
we apply them to financial and metabolic networks and find that inclusion of weights may 
considerably modify the conclusions obtained from the study of unweighted characteristics. 
\end{abstract}

% BEFORE: We introduce motif \emph{intensity} as the geometric mean of its link weights and 
% motif \emph{coherence} as the ratio of the geometric and corresponding arithmetic mean.
%
% This is not true. We define a motif as a set (ensemble) of topologically equivalent subgraphs of a network. 
% In the earlier version, ``its link weights'' could be understood to refer to all the weights in the ensemble, 
% whereas we really take the geometric average over subgraph instances and then sum these together. 

\maketitle

The network approach to complex systems has turned out to be extremely
fruitful and it has revealed some general principles applicable
to a large number of systems. Studies have produced 
unexpected findings such as the ubiquity of scale freeness, the
frequent appearance of high clustering, and the relationship between
functionality and the high appearance frequency of specific motifs. 
This approach has
also led to a number of novel paradigmatic models, providing a holistic
framework in which the details of the interactions between the
constituents of the complex systems are disregarded and only their
scaffolds are considered~\cite{network}.

A deeper understanding of these systems requires that, in
addition to the underlying network structure, information about the
strength of interactions is also taken into account. 
This is accomplished by assigning weights to the links, such as 
transportation fluxes in the Internet and air traffic networks~\cite{traffic,BBPV_PNAS}, 
or the reaction fluxes building the metabolic pathways of a cell \cite{metabolic}. 
Weights can also be obtained by applying a classification (or clustering) scheme to a correlation matrix, 
or for understanding the structure underlying the dynamics of microarray 
\cite{microarray} and stock market data~\cite{econo,okk_epj}.
Optimal paths~\cite{optimal} and minimum spanning trees~\cite{mst} 
also clearly depend on the distribution of weights.
These examples indicate the need to generalize the network characteristics
to weighted networks. Some recent efforts towards this goal are 
the discussion of the clustering coefficient for node weights 
\cite{node}, introduction of a  definition for the link weighted 
case~\cite{BBPV_PNAS,BBV}, and the mapping of weighted networks to 
multigraphs~\cite{multigraphs}. 
Our aim in this Letter is to introduce a set of practical tools that may be 
used to study the structure of a diverse group of systems where  
interactions strengths can be obtained and where omitting them would 
lead to a considerable loss of information. Many biological and 
social systems are expected to fall into this category.

% intensity
In general, we consider any weighted network as a fully 
connected graph where some of the links bear zero weights. 
For simplicity, we deal with (directed or 
undirected) networks where the weight $w_{ij}$ between 
nodes $i$ and $j$ is non-negative and not necessarily normalized.
We introduce the \emph{intensity} $I(g)$ of  
subgraph $g$ with vertices $v_g$ and links $\ell_g$ as the 
\emph{geometric mean} of its weights: 
\begin{equation} 
I(g)=\left(\prod_{(ij)\in \ell_g} w_{ij}\right) ^{1/|\ell _g|}, 
\label{eq:geom_mean} 
\end{equation} 
where $|\ell _g|$ is the number of links in $\ell_g$.
The definition suggest a shift in perspective from regarding subgraphs as discrete 
objects (either exist or not) to a continuum of subgraph intensities, where 
zero or very low intensity values imply that the subgraph in question 
does not exist or exists at a practically 
insignificant intensity level. In practice, low intensity values could result, 
for example, from measurement noise.

%coherence
Due to the nature of the geometric mean, the subgraph intensity $I(g)$
may be low because one of the weights is very low, or it may result 
from all of the weights being low. In order to distinguish between 
these two extremes, we introduce subgraph \emph{coherence} $Q(g)$ 
as the ratio of the geometric to the arithmetic mean of the weights as 
\begin{equation} 
Q(g) = I|\ell _g|/\sum_{(ij)\in \ell_g} w_{ij}. 
\label{eq:coherence} 
\end{equation} 
Here $ Q \in [0,1]$ and it is close to unity only if the subgraph weights do 
not differ much, i.e. are internally coherent. 

The concept of a motif was originally introduced to denote ``patterns 
of interconnections occurring in complex networks at numbers that 
are significantly higher than those in randomized networks'' 
\cite{motif}. However, this has led to some confusion, which partly 
stems from the specification of the random ensemble, i.e. the underlying 
null hypothesis \cite{motcrit}. 
We define a motif as a set (ensemble) 
of topologically equivalent subgraphs of a network. 
With weighted networks it becomes more natural to deal 
with intensities as opposed to numbers of occurrence, where the 
latter is obtained as a special case of the former. The motifs 
showing statistically significant deviation from some reference system 
can then be called high or low intensity motifs. 

\begin{figure}
%\vskip2mm
\includegraphics[width=7cm]{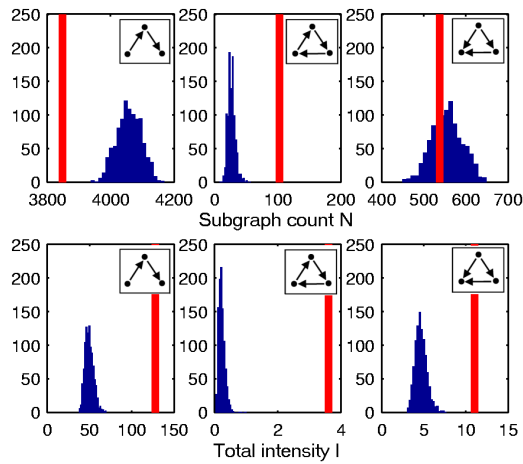}
\caption{Motif intensities for the empirical network (vertical lines) and
 the corresponding random ensembles (histograms), for the unweighted (upper panel) and
 weighted (lower panel) cases. }
\label{fig:motifs}
\end{figure}

% total intensity and total intensity in matrix form
We define the \emph{total intensity} $I_M$ of a motif $M$ in the network
as the sum of its subgraph intensities $I_M = \sum_{g \in M} I(g)$. 
For certain weighted directed motifs, the total intensities can be computed 
using simple matrix operations. Let the $N \times N$ weight matrix
$\mathbf{W}$ describe the network weights. Analogously, 
let $\mathbf{A}$ represent the underlying $N \times N$ adjacency matrix
such that $a_{ij}=1$ if $w_{ij}>0$, and $a_{ij}=0$ if $w_{ij}=0$.
In an unweighted network, the number of directed paths returning to
the starting node after $k$ steps can be written as

\begin{equation}
N(k) = \sum_{i_1, \ldots, i_k} \prod_{x=1}^{k} a_{i_{x},i_{x+1}} 
     =  %\sum_{i_1, \ldots, i_k} a_{i_1,i_2} a_{i_2,i_3} \cdots a_{i_k,i_1} =
        \textrm{Tr} \{ \mathbf{A}^k \}, 
\end{equation}

\noindent where the summation goes over all possible sites and
$i_{k+1}=i_{1}$ \cite{network}. Let $\mathbf{W}^{(1/k)}$
represent a matrix obtained from $\mathbf{W}=[w_{ij}]$ by taking
the $k$-th root of its individual elements such that
$\mathbf{W}^{(1/k)}=[w_{ij}^{1/k}]$. The total
intensity of motif $M$ in the network is
\begin{equation}
I_{M} = a_M \sum_{i_1, \ldots, i_k} \left( \prod_{x=1}^{k} w_{i_{x},i_{x+1}} \right) ^{1/k} \\
      = % \sum_{i_1, \ldots, i_k} w_{i_1,i_2}^{1/k} w_{i_2,i_3}^{1/k} \cdots w_{i_k,i_1}^{1/k} 
        a_M \textrm{Tr}\{\mathbf({\mathbf{W}}^{(1/k)})^k\}, 
\end{equation}
where $a_M$ is a combinatorial factor ensuring that each subgraph is 
counted only once. For example, for the non-frustrated triangle (Fig. \ref{fig:motifs}, middle column)
the total intensity becomes 
$I_{\Delta} = \frac{1}{3} \textrm{Tr} \{ (\mathbf{W}^{(1/3)})^3 \}$. 
A change in the direction of a link can be taken into account using the matrix transpose. 
For some motifs, such as the path of order 2 (Fig. \ref{fig:motifs}, left column) 
we need a ``block'' matrix $\mathbf{B} = [b_{ij}]$ to prevent us from double-counting subgraphs. 
In this matrix the diagonal 
elements $b_{ii}=0$ and for the non-diagonal elements $b_{ij}=0$ when $a_{ij}=1$  or $a_{ji} = 1$, 
and otherwise $b_{ij}=1$. This allows us to write the total intensity of the path motif as 
$I_{\angle} = \textrm{Tr} \{\mathbf{W}^{(1/2)} \mathbf{W}^{(1/2)}\mathbf{B} \}$. 
We prevent double counting here for reasons of compatibility with earlier work,  
but find that it poses no serious problem as long as the system of counting is systematically applied 
both in the empirical and random case. Double counting could, in fact, be desirable if the interaction strength 
measurements are noisy. Envision adding a small number $\epsilon$ to every 
link (including the zeros) to represent a noise component. Larger subgraphs may now simply consist of noise.

% z-scores
In \cite{motif} the $z$-score for studying the statistical significance of motif
occurrences was defined as 
\begin{equation} 
z_M=(N_M - \langle n _M \rangle)/\sigma _M, 
\label{eq:score} 
\end{equation} 
where $N_M$ is the number of subgraphs in motif $M$ in the empirical network 
and $\langle n_M\rangle$ is the expectation of their number in the 
reference ensemble, and $\sigma _M $ is the standard deviation 
of the latter. Replacing the number of subgraphs  
by their intensities generalizes the $z$-score to \emph{motif intensity score}
\begin{equation} 
\tilde z_M = (I_M - \langle i_M\rangle)/ ( \langle i^2_M\rangle - \langle i_M\rangle^2 )^{1/2},
\label{eq:weighted_score} 
\end{equation} 
where $i_M$ is the total intensity of motif $M$ in one realization of the 
reference system. It is clear that Eqs. \eqref{eq:score} and 
\eqref{eq:weighted_score}  
coincide for binary weights, implying that $\tilde z \to z$ in the limit.  
As an analogue to the motif intensity score, we 
introduce the \emph{motif coherence score} as  
\begin{equation} 
\widetilde {z'}_M = (Q_M - \langle q_M\rangle)/ ( \langle q^2_M\rangle-\langle q_M\rangle^2 )^{1/2}, 
\label{eq:weighted_score_coh} 
\end{equation} 
where $Q_M$ and $q_M$ are the total coherence for motif $M$ in the 
empirical network and in one realization of the reference system,
respectively. As the coherence of an unweighted subgraph is unity, also 
$\widetilde {z'} \to z$ as the weights become binary. 

% standard clustering coefficient
Triangles are among 
the simplest nontrivial motifs and they play an important role
as one of the basic quantities of network characterization in 
defining the \emph{clustering coefficient} $C_i$ at node $i$ as
\begin{equation} 
C_i= \frac{2t_i}{k_i(k_i-1)},   
\label {eq:clustering} 
\end{equation} 
where $k_i$ is the degree of node $i$ and $t_i$ is the number of 
triangles attached to the node \cite{network,SzAK}.
This quantity is normalized between 
0 and 1, and it characterizes the tendency of the nearest neighbors 
of node $i$ to be interconnected.

% intensity node averages and our cc
As triangles are one type of subgraph, the definition in 
Eq. \eqref{eq:geom_mean} may be used to yield the 
\emph{weighted clustering coefficient} $\widetilde{C_i}$
by replacing the number of triangles $t_i$ in Eq. 
\eqref{eq:clustering} with the sum of triangle intensities as 
\begin{equation}
\widetilde{C_i} = \frac{2}{k_i(k_i-1)} \sum _{j,k} (\tilde{w}_{i,j} \tilde{w}_{j,k} \tilde{w}_{k,i})^{1/3}, 
\label{eq:weighted_clustering}
\end{equation} 
where we use weights scaled by the largest weight in the 
network, $\tilde{w}_{ij} = w_{ij} / \max(w_{ij})$. 
This definition fulfills the requirement that $\widetilde{C_i} \to C_i$ 
as the weights become binary. We can relate the unweighted and weighted 
clustering coefficients through the \emph{average intensity} 
of triangles at node $i$ as $\bar I_i = \frac{1}{t_i} \sum_{g \in \mathcal{N}(v_i)} I(g)$, 
where $\mathcal{N}(v_i)$ denotes the neighborhood of node $i$, and this 
allows us to write the weighted clustering coefficient as 
\begin{equation} 
\widetilde {C_i} =\bar {I_i} C_i.
\label{eq:clustering_product} 
\end{equation} 
This equation gives a plausible interpretation of the weighted
clustering coefficient: It is the unweighted (topological) clustering coefficient 
renormalized by the average intensity. 
Naturally, a weighted clustering coefficient $\widetilde C'_i$ can also be 
formulated by renormalizing the unweighted coefficient by the average 
coherence $\bar{Q_i}$, instead of the average intensity $\bar{I_i}$, around node $i$.

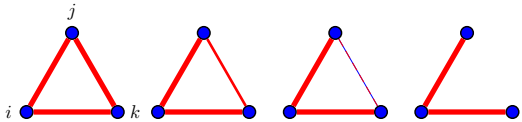
\begin{figure}
%\vskip2mm
\scalebox{0.70}{
\begin{pspicture}(-1,-1)(\textwidth, 3)
  \psset{nodesep=0.0, fillstyle=solid,fillcolor=blue}
  \psset{xunit=1.0cm, yunit=1.0cm}
  \rput(-0.5, -1.5){\scalebox{1.0 1.0}{
  \rput(2.000, 2.600){\circlenode{N3a_1}{}}
  \rput(2.000, 3.000){\psframebox[linestyle=none,fillstyle=none]{$j$}}
  \rput(1.134, 1.100){\circlenode{N3a_2}{}}
  \rput(0.800, 1.100){\psframebox[linestyle=none,fillstyle=none]{$i$}}
  \rput(2.866, 1.100){\circlenode{N3a_3}{}}
  \rput(3.200, 1.100){\psframebox[linestyle=none,fillstyle=none]{$k$}}
  \ncline[linecolor=red,linewidth=0.1cm]{N3a_1}{N3a_2}
  \ncline[linecolor=red,linewidth=0.1cm]{N3a_1}{N3a_3}
  \ncline[linecolor=red,linewidth=0.1cm]{N3a_2}{N3a_3}
  \rput(4.500, 2.600){\circlenode{N3b_1}{}}
  \rput(3.634, 1.100){\circlenode{N3b_2}{}}
  \rput(5.366, 1.100){\circlenode{N3b_3}{}}
  \ncline[linecolor=red,linewidth=0.1cm]{N3b_1}{N3b_2}
  \ncline[linecolor=red,linewidth=0.05cm]{N3b_1}{N3b_3}
  \ncline[linecolor=red,linewidth=0.1cm]{N3b_2}{N3b_3}
  \rput(7.000, 2.600){\circlenode{N3c_1}{}}
  \rput(6.134, 1.100){\circlenode{N3c_2}{}}
  \rput(7.866, 1.100){\circlenode{N3c_3}{}}
  \ncline[linecolor=red,linewidth=0.1cm]{N3c_1}{N3c_2}
  \ncline[linecolor=red,linewidth=0.01cm,linestyle=dashed]{N3c_1}{N3c_3}
  \ncline[linecolor=red,linewidth=0.1cm]{N3c_2}{N3c_3}
  \rput(9.500, 2.600){\circlenode{N3d_1}{}}
  \rput(8.634, 1.100){\circlenode{N3d_2}{}}
  \rput(10.366, 1.100){\circlenode{N3d_3}{}}
  \ncline[linecolor=red,linewidth=0.1cm]{N3d_1}{N3d_2}
  \ncline[linecolor=red,linewidth=0.1cm]{N3d_2}{N3d_3}
}}
\end{pspicture}
}
\caption{A schematic illustration of the difference between $\hat{C}_i$ and $\widetilde {C_i}$. 
The weight $w_{jk}$ is gradually decreased from left to right. The value of $\hat{C}_i$ 
is equal for the first three triangles and drops to zero suddenly for the fourth triangle as $w_{jk} \to 0$, implying that $a_{jk} = 0$. 
In contrast, the value of $\widetilde {C_i}$ decreases as $C_i \sim w_{jk}^{1/3}$, 
tending smoothly to zero in the limit.}
\label{fig:triangles}
\end{figure}

% BBV clustering
An alternative definition for weighted clustering coefficient was given in \cite{BBPV_PNAS} as 
\begin{equation} 
\hat{C}_i = \frac{1}{s_i(k_i-1)}\sum_{j,k} \frac{(w_{ij} + w_{ik})}{2} a_{ij} a_{ik} a_{jk}, 
\label{eq:BBV_clustering} 
\end{equation} 
where $s_i$ denotes the strength of node $i$, defined as $s_i = \sum_j w_{ij}$, and 
$a_{ij}$ is an element of the underlying binary adjacency matrix.
This definition considers only two of the three link weights, namely those adjacent to node $i$ 
($w_{ij}$ and $w_{ik}$) and requires that a link exist also between nodes $j$ and $k$ but does not 
take its weight ($w_{jk}$) into account. The difference between the two weighted clustering coefficients 
$\widetilde {C_i}$ and $\hat{C}_i$ is illustrated schematically in Fig. \ref{fig:triangles}.

% ----------------------------------------------------

Next we apply these concepts to two real networks.

\emph {(A) Undirected financial network.} We considered a set of 
daily price data for $N=477$ NYSE traded stocks 
from 1980 to 2000. We calculated the correlation matrix by extracting 4-year return windows 
in order to study the system's dynamics. Here the nodes correspond to stocks, and the weighted undirected links
to the elements in the correlation matrix. Thus, the stronger the weight, the stronger the coupling between 
the stock returns in terms of their linear correlation. The links are inserted in the network in descending order 
starting from the strongest one until a predetermined number of links has been reached. 
The method is described in detail in \cite{okk_epj}. 

\begin{figure}
\vskip2mm
\includegraphics[width=7.0cm]{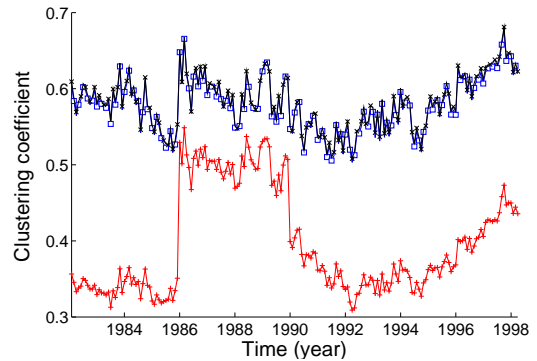}
\caption{Average clustering coefficients for the financial network.
  The weighted clustering coefficient $\widetilde C$ ($+$) of Eq. \eqref{eq:weighted_clustering}
  shows the effect of Black Monday clearly. The unweighted $C$ ($\Box$) of Eq. \eqref{eq:clustering} 
  and the weighted $\hat{C}$ ($\times$) of Eq. \eqref{eq:BBV_clustering} practically coincide 
  (the markers $\Box$ and $\times$ are used alternately).}
\label{fig:nyse}
\end{figure}

We have shown earlier that the famous Black Monday (10/19/1987) causes a temporary transition not only in the 
topology but also in the weights of the network \cite{black_monday}. 
Our aim is to use it as an example of a network undergoing this type of two-fold transition  (topology and weights) 
and to see whether the changes are reflected in the network's clustering statistics. 
In Fig. \ref{fig:nyse} we show the three clustering coefficients, averaged over the network, as functions of time:  
the unweighted $C$ of Eq. \eqref{eq:clustering}, 
the weighted $\hat{C}$ introduced in \cite{BBPV_PNAS} and given in Eq. \eqref{eq:BBV_clustering}, 
and the weighted $\widetilde C$ introduced in Eq. \eqref{eq:weighted_clustering}. 

The crash is not seen very clearly in $C$, as it can only capture the topological aspects 
of the transition. The weighted coefficient $\hat{C}$ is also fairly 
insensitive to the changes in link weights and practically coincides with $C$.  
The fact that $\widetilde{C}$ does reflect the transition indicates its ability to capture both aspects 
of the transition.  
The average values for the clustering coefficients outside (inside) the crash 'period'  
are $C=0.57$ ($C=0.60$), $\hat{C} = 0.58$ ($\hat{C} = 0.60$), and $\widetilde{C} = 0.36$ 
($\widetilde{C} = 0.50$). These numbers imply that $C$ and $\hat{C}$ 
increase less than 5\% during the crash which is less than their normal (outside the crash period) fluctuation, 
measured at 6.2\% as their standard deviation relative to the mean. 
However, the crash increases $\widetilde{C}$ by 39\%, which is considerably larger than 
the the level of fluctuation at 9.7\%. Thus, $\widetilde{C}$ has a considerably higher 
``signal-to-noise'' ratio. 
The results are not affected significantly by the value of the predetermined threshold. In the limit 
of inserting all the links of the correlation matrix,
we obtain a fully connected network for which
$C = \hat{C} = 1$ for all times, whereas $\widetilde C$ still shows the
effect of the crash clearly.

\emph{(B) Directed metabolic network.} Cellular metabolism can be represented as a
directed network of intracellular molecular interactions. The network
consists of nodes $X_{i}, Y_{j}$, which represent the chemicals
and they are linked if connected by a metabolic reaction. Here
we focus on the metabolic pathways of the bacterium
\emph{Escherichia coli} grown in glucose, which has
been studied intensely \cite{metabolic}.
In order to experiment with weighted directed motifs, we define the
weights through a biochemical reaction of the
form $x_{1}X_{1} + \cdots x_{n}X_{n} \to y_{1}Y_{1} + \cdots y_{m}Y_{m}$
with a positive (negative) net flux $f$ if the balance of the reaction lies to the right (left).
The flux provides an overall measure of the relative activity
of each reaction.
We define the weights as $w_{ij} = ({y_{j}}/{x_{i}}) f$, reflecting the rate
at which $X_{i}$ is converted into $Y_{j}$.

In order to employ motif intensity scores a reference system, corresponding to a null hypothesis, 
needs to be established. We follow a typical approach by constructing an ensemble of random 
networks by conserving the degree sequence of the empirical network using a switching algorithm \cite{motif}, 
which preserves the single-node characteristics of the empirical network. 
The weights are obtained simply by permuting the empirical weights. While removing any weight correlations,  
the approach guarantees conservation of the empirical weight distribution.

We summarize our findings in Fig.~\ref{fig:motifs}, in which we show
the unweighted and weighted motif intensities for a subset of the studied motifs:
(i) path of order 2, (ii) non-frustrated triangle, and (iii) frustrated triangle.
The motif intensity scores for the unweighted networks, which are based on the subgraph counts,
are $z_{i} = -5.4$, $z_{ii}=12.8$, and $z_{iii}=-0.5$, and for the weighted networks
$\tilde z_{i} = 14.8$, $\tilde z_{ii} = 33.8$, $\tilde z_{iii} = 9.0$.
These results show that a move from unweighted to weighted characteristics
can cause a change from low to high intensity, i.e. from under-representation to
over-representation. The intensity may become amplified, i.e. increase the extent of over-representation,
or it may increase from average to high intensity, i.e. from statistically
insignificant to over-representation.

%We summarize our findings in Fig.~\ref{fig:motifs} with weighted
%and unweighted motif intensity scores ($z$-scores).
%For motifs $a$ and $b$ we find no statistically significant
%deviation from the null hypothesis. For motif $c$ the unweighted
%$z$-score shows significant underrepresentation (low intensity),
%whereas for $\tilde z$ the motif is significantly overrepresented
%(high intensity). For the non-frustrated triangle $d$ both measures
%indicate over-representation, the effect being dramatically
%amplified for the weighted case. Finally, the frustrated triangle
%$e$ becomes significantly overrepresented under
%the weighted $z$-score but shows no deviation for $z$. These findings
%strongly support the incorporation of weights into the study of motif
%intensity scores.

%> (N_ang_1b - mean(N_ang_1b_r)) / std(N_ang_1b_r)   -5.4123
%> (N_ang_1w - mean(N_ang_1w_r)) / std(N_ang_1w_r)   14.8261
%> (N_ang_2b - mean(N_ang_2b_r)) / std(N_ang_2b_r)    0.5028
%> (N_ang_2w - mean(N_ang_2w_r)) / std(N_ang_2w_r)    1.5977
%> (N_ang_3b - mean(N_ang_3b_r)) / std(N_ang_3b_r)    0.5028
%> (N_ang_3w - mean(N_ang_3w_r)) / std(N_ang_3w_r)   -0.0901
%> (N_tri_1b - mean(N_tri_1b_r)) / std(N_tri_1b_r)   12.8227
%> (N_tri_1w - mean(N_tri_1w_r)) / std(N_tri_1w_r)   33.8152
%> (N_tri_2b - mean(N_tri_2b_r)) / std(N_tri_2b_r)   -0.5028
%> (N_tri_2w - mean(N_tri_2w_r)) / std(N_tri_2w_r)    8.9577

%-----------------------------------------------------------------------
In this Letter we have proposed two new concepts for
the characterization of weighted complex networks: the \emph{intensity} and \emph{coherence} of a subgraph.
They allow for a very natural generalization of the $z$-scores to
\emph{motif intensity scores} (Eqs. \ref{eq:weighted_score} and \ref{eq:weighted_score_coh}),
and the clustering coefficient to \emph{weighted clustering coefficient} (Eq. \ref{eq:clustering_product}).
Our studies with undirected financial networks show that
the weighted clustering coefficient reflects the effects of a market crash which 
is hardly observed with other clustering characteristics studied. 
Our results on the directed metabolic network of E. Coli
indicate that incorporation of weights into network motifs may
considerably modify the conclusions drawn from their statistics.

%Using these concepts a natural generalization
%of the clustering coefficient emerges (Eq. \ref{eq:clustering_product})
%and the weighted score of motifs was introduced
%(Eqs. \ref{eq:weighted_score} and \ref{eq:weighted_score_coh}). We applied
%these characteristics to three different weighted networks: the
%network produced by the BBV model, the network based on
%correlations between stock returns for a set of NYSE stocks, and the (directed)
%metabolic network of E. Coli. Our results indicate that the
%incorporation of weights into the study of network motifs may
%considerably modify the conclusions drawn from their statistics.
%This approach could be used to obtain deeper insight into the relationship
%between network structure and functionality.
 
Acknowledgments: We are thankful to A.-L. Barab\'asi, E. Almaas and 
S. Wuchty for the metabolic network data and useful discussions. 
This work was carried out at the Center of Excellence of the Finnish 
Academy of Sciences, Computational Engineering, HUT. JK is partially 
supported by the Center for Applied Mathematics and Computational Physics, BUTE.

\end{document}